\documentclass{iau}
\usepackage{graphicx,amsmath}

\title[Mass transport in V346~Nor] 
      {Envelope-to-disk mass transport in the FUor-type young eruptive star V346 Normae}

\author[\'A.~K\'osp\'al et al.]  
       {\'A. K\'osp\'al$^{1,2}$, P. \'Abrah\'am$^1$, O. Feh\'er$^1$,
         F. Cruz-Saenz de Miera$^1$, M.~Takami$^3$}

\affiliation{$^1$Konkoly Observatory, Research Centre for Astronomy
  and Earth Sciences, Hungarian Academy of Sciences, Konkoly-Thege
  Mikl\'os \'ut 15-17, H-1121, Budapest, Hungary \\ email: {\tt
    kospal@konkoly.hu} \\[\affilskip] $^2$Max Planck Institute for
  Astronomy, K\"onigstuhl 17, D-69117 Heidelberg, Germany \\[\affilskip]
  $^3$Institute of Astronomy and Astrophysics, Academia Sinica, PO Box
  23-141, 106, Taipei, Taiwan}

\pubyear{2019}
\volume{345}  
\jname{Origins: from the Protosun to the First Steps of Life}
\editors{Bruce G. Elmegreen, L. Viktor T\'oth, Manuel G\"udel, eds.}

\begin{document}

\maketitle

\begin{abstract}
Having disk-to-star accretion rates on the order of
10$^{-4}\,M_{\odot}$/yr, FU Orionis-type stars (FUors) are thought to
be the visible examples for episodic accretion. FUors are often
surrounded by massive envelopes, which replenish the disk material and
enable the disk to produce accretion outbursts. We observed the
FUor-type star V346~Nor with ALMA at 1.3~mm continuum and in different
CO rotational lines. We mapped the density and velocity structure of
its envelope and analyzed the results using channel maps,
position-velocity diagrams, and spectro-astrometric methods. We
discovered a pseudo-disk and a Keplerian disk around a
0.1\,$M_{\odot}$ central star. We determined an infall rate from the
envelope onto the disk of 6$\times$10$^{-6}\,M_{\odot}$/yr, a factor
of few higher than the quiescent accretion rate from the disk onto the
star. This hints for a mismatch between the infall and accretion rates as
the cause of the eruption.  
\end{abstract}


A long-standing open issue of the paradigm of low-mass star formation
is that most protostars are less luminous than theoretically
predicted.  One possible solution is that the accretion process is
episodic. The young outbursting FU Ori-type stars (FUors) are thought
to be the visible examples for objects in the high accretion state.
Mass infall from FUor envelopes may play a fundamental role in
refilling the disk, and triggering instabilities causing accretion
outbursts. However, our knowledge on the envelope dynamics and on the
envelope-to-disk mass transfer is still very limited.

V346~Nor is an embedded FUor in the Sandqvist~187 dark cloud, at a
distance of 700\,pc. It erupted between 1976 and 1980, stayed in the
bright state until 2008, rapidly faded in 2010, and is now showing a
partial re-brightening. We modeled the multi-epoch optical-infrared
spectral energy distributions and found a peak accretion rate of
10$^{-4}\,M_{\odot}$/yr in 1992, and a drop of accretion by at least a
factor of 100 in 2010, suggesting that the quiescent accretion rate is
less than about $10^{-7} - 10^{-6}\,M_{\odot}$/yr (\cite{kospal2017a}).

We observed V346~Nor with ALMA in the J = 2--1 line of $^{12}$CO,
$^{13}$CO, C$^{18}$O, and 1.32\,mm continuum in one single setting in
Band 6. We used a combination of 12\,m array configurations, as well
as data from the 7\,m array and Total Power antennas. We had two runs:
in 2014--15 the beam size was 1$\overset{\prime\prime}{.}0$ with a
continuum sensitivity of 70 $\mu$Jy/beam, while in 2017--18 the beam
size was 0$\overset{\prime\prime}{.}$1 with a continuum sensitivity of
35 $\mu$Jy/beam.

\begin{figure}[b]
\vspace*{-.1 cm}
\begin{center}
 \includegraphics[width=5.1in]{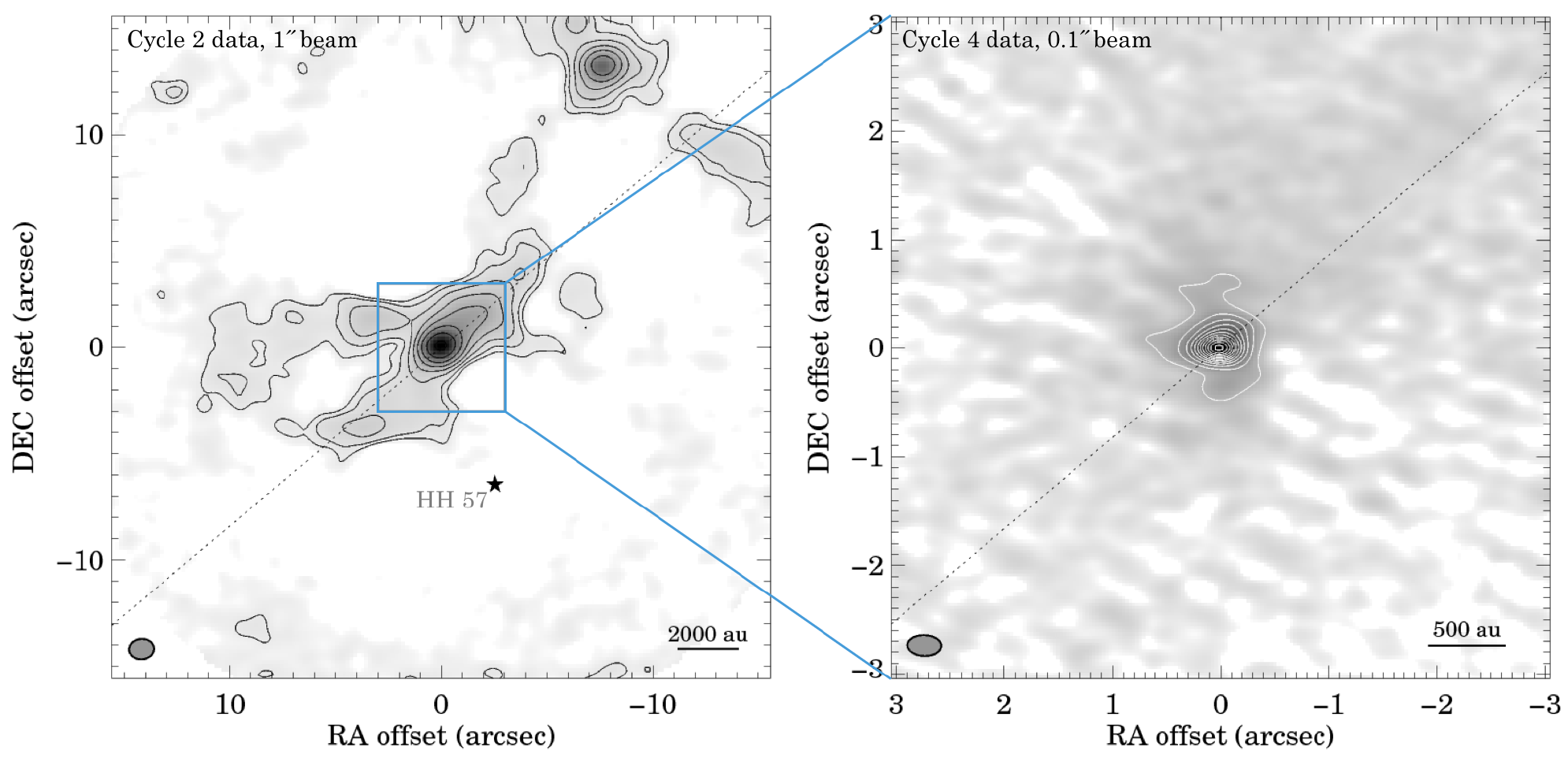}
\vspace*{-.1 cm}
 \caption{ALMA 1.32\,mm continuum map of V346~Nor (left panel
   from \cite{kospal2017b}).}
   \label{fig:alma}
\end{center}
\end{figure}

We detected a fairly compact continuum source coinciding with the
near-infrared stellar location (Fig.~\ref{fig:alma}). The source is
marginally resolved, with a deconvolved FWHM of
0$\overset{\prime\prime}{.}$21$\times$0$\overset{\prime\prime}{.}$15. We
interpret this as an inclined disk-like object, with a radius of
50\,au. The position angle is NW-SE, same as the CO rotating
disk. The central source is surrounded by more extended, fainter
emission out to 6300\,au. The dust mass in the compact source within
50\,au is 3$\times$10$^{-4}\,M_{\odot}$.

Results from our spectro-astrometric analysis of the C$^{18}$O data
cube indicates that from 350 to 700\,au, the radial velocity profile
is consistent with a pseudo-disk (infalling-rotating motion with
angular momentum conservation). The inner 350\,au resembles a
Keplerian disk around a 0.1$\,M_{\odot}$ central star. The total gas
mass within 350/700\,au is 0.01/0.03$\,M_{\odot}$.

Both the $^{12}$CO and $^{13}$CO profiles exhibit high-velocity line
wings. The redshifted emission shows a parabola opening towards the
northeast with a relatively wide opening angle of about
80$^{\circ}$. The blueshifted emission forms a narrower ellipse
extending towards the southwest with an opening angle of about
40$^{\circ}$. The geometry is reminiscent of an outflow cavity, where
emission is coming from the swept-up material in the cavity walls. The
Herbig-Haro object HH~57 is situated along the axis of the
southwestern CO-emitting ellipse, close to its farther edge.

We calculated the infall rate using an analytic formula from
\cite{momose1998}. The resulting infall rate from the envelope onto
the disk is 6$\times$10$^{-6}\,M_{\odot}$/yr, which is a factor of few
higher than the quiescent accretion rate from the disk onto the
star. It is a hint for a mismatch between the infall and accretion
rates. This is the first observational support for such mismatch in a
FUor, previously invoked to explain FUor outbursts.

Acknowledgements. This project has received funding from the
European Research Council (ERC) under the European Union's Horizon
2020 research and innovation programme under grant agreement No 716155
(SACCRED, PI: \'A. K\'osp\'al). This paper makes use of the following ALMA
data: ADS/JAO.ALMA\#2013.1.00870.S, ADS/JAO.ALMA\#\\2016.1.00209.S. ALMA
is a partnership of ESO (representing its member states), NSF (USA)
and NINS (Japan), together with NRC (Canada) and NSC and ASIAA
(Taiwan) and KASI (Republic of Korea), in cooperation with the
Republic of Chile. The Joint ALMA Observatory is operated by ESO,
AUI/NRAO and NAOJ.

\end{document}